# From the Big Bang to *Island Universe*: Anatomy of a Collaboration[1]

By David H. Weinberg, Ohio State University Department of Astronomy, June 2010

I first met Josiah McElheny in September 2004, in the café of the Wexner Center for the Arts, for what turned out to be a 3-hour lunch. We were joined by Helen Molesworth, the Wexner Center's head curator. From an earlier conversation with Helen, I knew that Josiah wanted to make a sculpture about the big bang, connected in some way to the chandeliers that hang in New York's Metropolitan Opera House. Conveying an accurate understanding of the big bang is one of the hardest challenges I face in my introductory astronomy lectures, and I doubted that I could provide useful advice on how to do it with a sculpture. But by the end of our conversation we had identified a way forward, and after six years and four completed works, Josiah and I are still collaborating.

Josiah and I have given numerous talks in the last two years, some individually and some jointly, on the four sculptures that have so far emerged from our collaboration: *An End to Modernity* (2005), *The Last Scattering Surface* (2006), *The End of the Dark Ages* (2008), and *Island Universe* (2008). I was delighted to participate in the *Narrative, Science, and Performance* symposium, which was an especially apt forum because of our project's deep roots at Ohio State and the Wexner Center, and because of the symposium's focus on the nexus between science and art. I have written elsewhere (see the Bibliography) about the scientific background of these works, explaining the "code" that connects their elaborate arrangements of chrome-plated metal, glass disks and globes, and incandescent lamps to astronomical theories and observations. Here I will give just brief explanations of the science and focus instead on the history of the project, providing what I hope is an informative "case study" of an ambitious collaborative effort at the art/science border.

### *An End to Modernity*

In that first lunchtime conversation, Josiah began by showing me examples of his recent work, intricate constructions of transparent and reflective glass that explored themes of modernist design and architecture. He described his fascination with the Met chandeliers, a striking hybrid of cut-glass opulence and Sputnik spikiness, created in 1965, the year that the discovery of the cosmic microwave background provided the linchpin evidence for the big bang theory. Josiah showed me a sketch of the sculpture he envisioned: a "Met chandelier" remade in a streamlined modernist style, brought down to eye level, and connected explicitly to the big bang, with glass pieces representing galaxies and lamps representing quasars.

Josiah knew the potential conceptual flaw of this design. While the popular image of the big bang is an explosion of material from a central point into pre-existing space -- an atom bomb on steroids -- cosmologists understand the big bang as the origin of space and time itself, initiating an expansion that occurs everywhere and has no center. I ran through my usual analogy of the expanding universe as the skin of an inflating balloon, carrying apart "galaxies" marked on its surface. Returning to Josiah's sketch, I noted that one might naturally read it not as a snapshot of a single instant but as a progression in time, with explosion at the center leading to subsequent fragmentation in the outer regions. I suggested that we make this progression explicit by having distance from the center represent time, mapping specific locations in the sculpture to specific epochs of cosmic history. The outer edge would mark the present day, and the supporting central sphere could represent the "last scattering surface," formed half a million years after the big bang, when the opaque fog of light-scattering free electrons dissipated and the universe became transparent for the first time. Somewhat to my surprise, Josiah embraced these suggestions immediately. And really, everything else has followed from there.

---

[1] To appear in the journal *Narrative*, as part of the proceedings of the symposium *Narrative, Science, and Performance,* edited by J. Phelan. The symposium took place at Ohio State University in October, 2009, hosted by OSU's Project Narrative and the Wexner Center for the Arts.

Over the ensuing six months, Josiah made monthly visits to Columbus, and I gave him a crash course in modern cosmology, covering the empirical and theoretical underpinnings of the big bang theory and more recent discoveries about the formation and growth of galaxies and quasars. Each month Josiah would go back to New York with a stack of articles or books, and the next month he would return with a page of handwritten questions, which consistently went to the heart of the scientific or conceptual issues at stake. As our ideas about the structure of the sculpture took shape, we began to worry about technical questions. Rods emanating from the central sphere would terminate in either a lamp, representing a quasar, or a cluster of glass pieces, representing a cluster or supercluster of galaxies. How thick did the rods need to be so that they would not bend perceptibly under the weight of a cluster? Should they be uniform thickness or telescoped? How closely could the rods be spaced on the central sphere? Would the minimum spacing make the pattern of rods too regular, erasing the randomness that we hoped to achieve?

I wrote some simple computer scripts to show what the structure might look like for different choices about the number and sizes of the elements. As we expanded our explorations, these scripts gradually grew into a complex web of programs, which transformed the statistical rules that Josiah and I developed into a list of parts and assembly instructions. Along the way, these programs solved some of the more subtle technical problems, ensuring that "galaxies" would not overlap in space, that each cluster could be screwed onto its supporting rod without running afoul of a neighboring rod, and that the sculpture would not tip when suspended from the apex of its central sphere. Josiah, meanwhile, set about a far more daunting task, figuring out how to construct the ferociously complicated object that was emerging from our discussions and diagrams. In the end, the metal parts were fabricated to design by a company in California, and Josiah and his assistant Anders Rydstedt fabricated the 932 glass pieces in Josiah's studio over the course of a hot New York summer. In a frenzied final two weeks, Josiah and a team from the Wexner Center assembled the 2000+ parts, some of them still arriving in daily batches, into the finished sculpture.

*An End to Modernity* is 172'' across and 138'' tall, suspended so that the 28'' reflective central sphere hangs at eye level and the lowest glass pieces lie six inches above the floor. With our adopted mapping between location and time, the shortest, 36'' rods end 100 million years after the big bang. They support either faint lamps or groups of a few small glass pieces with the disk-like form expected for the earliest galaxies. Moving outwards towards the present, the glass pieces become larger, tracing the growth of galaxies as they attract gas from their surroundings and process it into stars. Glass spheres, representing elliptical galaxies formed by the chaotic collisions of stellar disks, become increasingly common, especially in the dense central regions of the clusters. The clusters themselves grow in size and complexity, acquiring the extended filamentary structure seen in 3-dimensional maps of the distribution of galaxies. Lamps become more numerous and brighter, then fade in the outermost zone, tracing the observed rise and fall of the quasar population.

*An End to Modernity* appeared first in Helen Molesworth's *Part Object/Part Sculpture* exhibition, an exploration of the varied legacies of Marcel Duchamp. Josiah chose the title about a month before the exhibition, and it came as a great surprise to me. It invariably puzzles my astronomical colleagues as well, so I have had to become reasonably proficient at explaining it. The sculpture plays on the historical and intellectual coincidence between the discovery of the cosmic microwave background and the design of the Met chandeliers, a "fictitious history" that Josiah brings out more explicitly in his companion film *Conceptual Drawings for a Chandelier (1965)*, which interleaves footage of the chandeliers with abstract diagrams drawn from astronomy and cosmology textbooks. Josiah also identifies 1965 as a time when developments in art and culture were "laying waste to the modernist view of history as a single line of progressive thought," when "intellectual thought in the West was beginning to splinter in a way that echoed cosmology's concept of a de-centered, non-hierarchical universe."[2] Of the many scientific issues we dug into while designing the sculpture, the one that fascinated Josiah the most was the tension between the homogeneity and isotropy of the universe on large scale and the specificity of individual structure on small scale, the notion that "any part of the universe is as likely to be 'boring' or 'interesting' as any other."[3] *An End to Modernity* -- and, by retrospective implication, the Met chandeliers themselves -- embodies the explosion of the unifying modernist ethos into the diverse fragments of the post-modernist era.

### *The Last Scattering Surface and The End of the Dark Ages*

The idea of making multiple sculptures was born of necessity. Because of its cost, *An End to Modernity* required substantial investments from Josiah's galleries in New York and Chicago, and additional pieces were needed to make the whole project financially viable. We considered the simple option, alternative realizations of *An End to Modernity* in which every detail would be different but the overall structure would be the same. We quickly decided, however, that we should embrace the opportunity to highlight other aspects of cosmology, exploring themes that we had discussed at length but had not tried to incorporate into the first sculpture.

---

[2] Josiah McElheny in *The Development and Origins of Island Universe*, which appears in the two *Island Universe* catalogs discussed in the Bibliography.
[3] *Ibid.*

In *An End to Modernity,* the last scattering surface is represented by the central reflective sphere. In reality, the last scattering surface is not a simple opaque barrier but the glowing fireball of the early universe, surrounding us at a distance of 14 billion light years. For the second sculpture, we decided to represent this glow by studding the central sphere with 150 incandescent lamps. The cosmic microwave background – the observed glow of the last scattering surface – is almost perfectly smooth, but its tiny, 0.001-percent variations are the seeds that gravity has amplified into the galaxies and large scale structure that we see today. We used varying lamp brightnesses to display (at a highly exaggerated level) these primeval microwave background fluctuations, as they were first mapped in the early 1990s by the Cosmic Background Explorer satellite. While *An End to Modernity* depicts the full history of galaxy and quasar evolution, *The Last Scattering Surface* focuses specifically on the relation between the primeval fluctuations and present day structure, so all of its rods are of a single length and all of its clusters resemble those at the outer edge of *An End to Modernity.*

As we were designing *The Last Scattering Surface*, we discussed it largely in practical terms, concentrating on how to solve the new technical challenges it created. Josiah was captivated, however, by the idea that the *imperfections* of the early universe were crucial to the development of structure, form, and life, and that a perfectly smooth universe would have remained perfectly sterile, forever. We brought out this theme more explicitly when we began to give joint talks about the sculptures in 2008, especially in our talk at the Phoenix Art Museum, where *The Last Scattering Surface* has found its permanent home in the museum's glass-walled entry lobby. There we posed *The Last Scattering Surface* as a counterpoint to Josiah's 2001 installation *Kärtner Bar, Vienna, 1908, Adolf Loos (White)*, in which pure white reproductions of Loos's designs and a white-on-white reprinting of Loos's essay *Ornament and Crime* show the modernist ideal of purity and perfection taken to its seductive but horrifying extreme.

*The End of the Dark Ages,* the third product of our collaboration, zeroes in on the first one billion years of cosmic history, when the diffuse glow of the big bang faded and the first stars, galaxies, and quasars filled the universe with discrete sources of visible light. In structural terms, it is like the inner regions of *An End to Modernity*, expanded in scale and detail to allow closer inspection. Astronomers define the end of the cosmic "dark ages" as the time when radiation from hot stars and young quasars broke the hydrogen atoms in intergalactic space back into protons and electrons, once again filling the universe with a free-electron fog. *The End of the Dark Ages* terminates at roughly this epoch, after revealing the emergence of vitality from primordial imperfection.

### *Island Universe*

Soon after Josiah imparted his plan to make multiple sculptures, I remarked that we could hang them all in one room and call the exhibit "Eternal Inflation." The inflation model, introduced by Alan Guth in 1980, proposes that the universe became homogeneous through an early phase of extremely rapid exponential expansion, followed by an epoch of "reheating" that produced the hot, radiation-filled cosmos of the conventional big bang theory. Inflation extends rather than replaces the big bang theory, explaining features of the universe that the standard model took for granted. In 1982, Andrei Linde suggested a variant of the inflation hypothesis in which an "inflationary sea" expands exponentially forever but constantly spawns "bubbles" within which reheating occurs and inflation ends. The inhabitants of a bubble see a "normal" big bang universe, but their entire observable cosmos is just one of many (perhaps infinitely many) denizens of the "multiverse." Each one of these cosmic bubbles could have different matter and energy contents or different initial fluctuation levels, or even -- the truly exotic twist -- different laws of fundamental physics or different numbers of spatial dimensions. Wildly speculative though it may sound, the multiverse inflation scenario is now widely accepted as a plausible account of the cosmos on scales far larger than those we can observe directly.

I gave Josiah a copy of Linde's 1987 article "Particle Physics and Inflationary Cosmology," which was my own introduction to eternal inflation. "P.P.I.C." appeared in *Physics Today*, which is a "popular" magazine in the sense of being comprehensible to the average Ph.D. physicist. By then I knew that Josiah could extract the key points from the introductions and conclusions of technical articles without getting lost in the mathematical middle. I also guessed, correctly, that he would like Linde's evocative, hand-drawn illustrations of the multiverse and its inhabitants. I had floated the idea of a multiverse installation as half joke, half provocation. Josiah was still wrestling with the technical and organizational challenges of the first sculpture, and the notion of building many of them to make a single work seemed beyond practical imagination. From his continuing questions about the Linde article, however, I could tell that the idea had gotten under his skin.

Some time later, I suggested that an eternal inflation piece could be titled "Island Universe," a term that I attributed (incorrectly, as we later learned) to Immanuel Kant. In a 1755 monograph titled *Universal History and Theory of the Heavens*, Kant proposed (correctly) that the nebulous elliptical smudges seen through telescopes were actually enormous, and enormously distant, stellar systems like our own Milky Way. A century later, Alexander von Humboldt popularized Kant's conjecture as the "island universe" hypothesis, but it was not until 1923 that Edwin Hubble demonstrated its validity, showing that the *closest* spiral nebula, Andromeda, was ten times further from earth than even the most distant stars of the Milky Way. Kant's hypothesis expanded the size of the known universe far beyond the visible stars, and indeed he concluded that it went on forever, filled with galaxies in all directions. Linde's hypothesis, I suggested, was analogous to Kant's, but on a scale vastly larger still --- infinities embedded in yet larger infinities. The cosmic islands of Kant's universe are separated by the practical impossibilities of intergalactic voyages, but the islands of Linde's multiverse are separated by the physical impossibility of traveling faster than light. As Josiah and I worked on *An End to Modernity* and *The Last Scattering Surface*, we would occasionally digress into discussions of inflation and the multiverse, Linde and Kant. When Josiah got an invitation from the White Cube Gallery in London to make "something that seemed impossible," he decided that *Island Universe* would be the project.

For me, the most memorable phase of *Island Universe* was Josiah's two-day visit to the Institute for Advanced Study, where I spent the fall of 2006. By the afternoon of the second day, we had covered a wall-to-wall blackboard with diagrams and notes on the five universes we had decided to represent. We came up with a working title for each universe: "Heliocentric" for the one that obeyed rules similar to *An End to Modernity*'s and that could therefore represent our own universe; “Small Scale Violence” for the one in which strong primordial fluctuations produced intense early formation of galaxies and quasars; “Frozen Structure” for the one in which a high fraction of repulsive dark energy arrested the growth of galaxies; “Late Emergence” for the one in which weak primordial fluctuations delayed galaxy formation; and “Directional Structure” for the one in which galaxies and quasars formed only in the disk-like plane of a “super-galaxy.” Some of these names were clunkier than others, but they quickly became indispensable as we tried to keep track of our evolving multiverse. All of them have stuck.

Attending the exhibition openings for all of these cosmological sculptures has been enormous fun – seeing the ideas and the parts lists and the computer plots embodied as real objects for the first time, discovering reflections and refractions that we did not anticipate, and meeting Josiah’s remarkable circle of friends. The opening of *Island Universe,* in London, was the best of all. Our sense of completion and celebration was enormous. The White Cube Gallery seemed like a perfect setting for the piece, an abstract, almost mathematical space whose gridded ceiling reflected from sphere to sphere. To my great regret, however, I missed the second exhibition of *Island Universe*, which was in a nearly antithetical setting that seems to have worked even better: the Reina Sofia Museum’s Palacio de Cristal, located in a park in central Madrid. There -- to judge from the photographs and from Josiah’s description -- the *Island Universe* sculptures looked like space-age objects that had materialized inside a 19$^{th}$-century building designed, somehow, with the knowledge that they would one day appear. Viewers wandered in from the park to encounter these enigmatic forms, whose appearance shifted slowly with the time of day and swiftly with passing clouds.

The two *Island Universe* exhibitions also gave rise to two striking catalogs, with exquisite images and an impressive array of texts. They include facsimiles of Linde’s article and of the crucial (translated) chapter of Kant’s *Universal History*, and a marvelous essay by Stanford philosopher Thomas Ryckman, which describes the history of Kant’s monograph and identifies startling parallels between Kant’s ideas and the motivations and implications of multiverse cosmology. I had suggested the analogy between Kant and Linde largely on a whim, knowing that it would catch Josiah’s attention, but the connections go much deeper than I could have guessed.

### *Conversations and Gifts*

From the beginning, our collaboration was grounded in conversation, which would often veer from minute technical details to long digressions on science, art, philosophy, fiction, or history. Sometimes these digressions would float in the air for months, then return suddenly to assume an important thematic role in one of the sculptures. The London exhibition of *Island Universe* provided an opportunity to take our conversation public, with a joint lecture at the Institute of Contemporary Arts. Over the next year, Josiah and I gave additional joint lectures at the annual meeting of the American Society for Aesthetics, at the Phoenix Art Museum, and at the Wexner Center (jointly sponsored, in a cross-disciplinary first, with Ohio State’s Center for Cosmology and Astro-Particle Physics). After each lecture, we would continue one-on-one, usually over dinner, each probing what the other had said, asking questions and proposing variations. Before the next lecture we would incorporate ideas or comparisons or transitions that had emerged from these discussions. The process of speaking about and writing about the sculptures has allowed both of us to understand them more deeply – philosophically, aesthetically, and scientifically -- than we did when we were working to create them.

Over the years, our private and public conversations have explored the differences between scientific models and artistic models, the color of the universe, the seductions and perils of modernism, the mix of philosophical preconception, mathematics, and language in scientific arguments, the minor role of reflected light in astronomically observable objects, the musical key of acoustic oscillations in the early universe, the chemistry of glass furnaces, and the multi-scale transitions between inhomogeneity and homogeneity in eternal inflation models. We have also developed the habit of giving gifts. Most of them are small – books, meals, a product-labeled can of "dark matter" that Josiah found at one of those places selling gag gifts for science geeks. Many of them are free – evocative quotations, or suggestions of articles that might interest or amuse. I have certainly profited the most from these exchanges, since Josiah can actually *make* things – glasses emblazoned with my daughter's initials, hand blown during two separate visits to the glass studio when she was four and eight, and, the biggest gift by far, a half-dozen glass clusters cloned from *An End to Modernity* and *Island Universe*, which hang in my house as elegant stand-alone sculptures and reminders of their larger kin.

My own most inspired gifts have been the Linde article, a quotation from Blaise Pascal by way of Jorge Luis Borges[4], and an optical fiber plugplate from the Sloan Digital Sky Survey. The Pascal quote – "the universe is an infinite circle, whose center is everywhere and whose circumference is nowhere" – struck me as a poetic metaphor for the isotropic universe and its centerless expansion. The plugplate is a 3-foot brushed aluminum disk, precision-drilled with 640 small holes in an irregular pattern and hand-labeled in black permanent marker, one of 2000 such plates used in creating the largest ever three-dimensional map of the universe. I sent it to Josiah without comment, but he figured out what it was immediately and hung it on the wall of his studio for contemplation.

Like the Linde article, the plugplate was intended partly as provocation, though of just what I wasn't sure. Now, following a long exhalation after *Island Universe*, we find ourselves collaborating on a new project, a very different kind of "chandelier" in which glass pieces and lamps suspended from a Sloan plate will show not an entire model universe but a core-sample of stars, galaxies, and quasars found along a specific line of sight through our own cosmos, from our own earthbound perspective. The Pascal quote, too, has returned. Following a completely different chain of associations, Josiah arrived at the 1872 book *Eternity by the Stars*, by the French communist leader Auguste Blanqui, who mysteriously chose to devote his imprisonment to writing a scientific tract instead of a political manifesto. Blanqui quotes Pascal in the second paragraph of the book, correcting "infinite circle" to "infinite sphere," and he goes on to a philosophically minded account of astronomical knowledge and to prescient speculations on the implications of infinity. Today, our continuing conversation zeroes in on the relative merits of 18mm vs. 24mm glass pieces and the difficulties of chrome-plating aluminum, then zooms out to the connections among Blanqui's speculations, Friedrich Nietzsche's theory of the Eternal Return, the combinatorics of Borges' short story *The Library of Babel*, and contemporary cosmological discussions of replicas and "Boltzmann brains" in an infinite universe.

---

[4] J. L. Borges, *The Fearful Sphere of Pascal*, in *Labyrinths*.

### *Lessons*

In the galleries, the sculptures appeared with a minimum of explanatory text – the wall plaques might mention cosmology or the big bang, but to learn the science behind the design you had to track down the catalog essay or hear me or Josiah explain it. This absence of information ran counter to my pedagogic impulse, and I was initially surprised that Josiah cared so much about getting accurate scientific details into the sculptures but wasn't bothered if viewers encountered the sculptures without a "key" to decode them. However, I have gradually come to share his point of view, that anyone who spends a few minutes with one of the sculptures can discern a complex system of ordered randomness behind it, a system that invites exploration and repays close inspection with deeper complexity. The viewer develops her own interpretation of this hidden information much like a scientist encountering a new natural phenomenon, and she can choose (or not) to compare this interpretation to the "back of the book" solution in the catalog essay. The different emphasis on explanation vs. discovery, I have concluded, is much of the difference between an exhibit for a science museum and an art work for a gallery.

What else can we learn from this "case study" about collaborations at the art/science border? First, it helps a great deal if each party in the collaboration has interest in and sympathy for the way the other works. Josiah clearly enjoyed learning about astronomy, not just the results but the nuts and bolts of how it is done. For my part, I have learned about modern and contemporary art from my wife, an art historian, and from her circle of friends, so I entered the project with an appreciation of what Josiah was trying to accomplish and excitement about making a contribution. Second, and equally important, the collaborators must find design principles that allow scientific and aesthetic ideas to reinforce each other rather than tug in opposite directions. In our case, that principle was the space-to-time translation that we mapped out in our very first conversation. Once we adopted that structure, every scientific detail we added to the sculpture made it aesthetically richer.

What's in it for the artist? For a start, science is one of the most influential elements of modern culture and human intellectual history, making it a natural subject for critical inquiry by other fields. In addition, science can be a powerful generator of visual ideas, woven with tensions and surprises and layers of interaction. After 13 months of work, *An End to Modernity* looked similar in general structure to the sketch Josiah showed me at the outset, but the details made all the difference, not just intellectually but visually. In Josiah's words: "The rules we developed to represent the actual evolving universe produced forms that were stranger, and much more complex and compelling than I could have come up with by mere invention."[5]

---

[5] *Ibid.*

What's in it for the scientist? First, it's a chance to reach far beyond one's usual audience, touching people whose lives and interests and outlook may be completely different from one's own. More people saw *Island Universe* in one day in Madrid than have ever read my Astrophysical Journal articles. The sculptures have been a great basis for popular talks in science venues, and it has been even more interesting to talk about them in art museums, where they elicit a wide range of reactions and questions. Second, it's a chance to learn from the inside about a radically different field of intellectual endeavor, encountering both the conceptual tools and the practical constraints of the arts. Seeing a set of ideas turn into a concrete, magnificent object is very rewarding. Lastly, it's a chance to indulge in the romantic aspect of science, which is something we tend to suppress in our professional work, but which has everything to do with becoming a scientist in the first place.

A final lesson from our experience is that Josiah and I understand far more about the sculptures in retrospect than we did at the time we conceived and designed them, as a result of talking to each other, speaking to audiences, writing, reading, and reflecting. Many connections that we first saw with vague intuition proved far richer than we could have hoped. So my advice to artists and scientists contemplating such a collaboration is to jump in, and see where it takes you, without trying to figure everything out ahead of time. In the end (if there is an end), it will be much more work than you bargained for, and much more fun than you imagined.

***Bibliography***

The best place to learn about Josiah's art in general and the cosmological sculptures in particular is the recently published book *Josiah McElheny: A Prism*, 250 pages of alternating images and texts, edited by Louise Neri and Josiah McElheny, and published by Skira/Rizzoli books (New York, 2010). The texts include my essay describing the scientific background to the four cosmological sculptures and several other essays that discuss these works in the broader context of Josiah's *oeuvre* and the history of modernism.

There are two dazzling catalogs of the two *Island Universe* exhibitions. The first, titled *Josiah McElheny – Island Universe*, is published by Jay Jopling / White Cube (London, 2009). The second, titled *A Space for an Island Universe*, is published by the Museo Nacional Centro de Arte Reina Sofia (Madrid 2009). Both of them include an edited "conversation" between Josiah and me on the development and origins of *Island Universe*, an essay on *Island Universe* by art historian Molly Nesbit, a translation of Kant's *Concerning Creation in the Total Extent of Its Infinity Both in Space and Time*, from his *Universal Natural History and Theory of the Heavens,* and an essay by philosopher Thomas Ryckman about Kant's monograph and its connections to *Island Universe.* In addition, the London catalog includes a facsimile reprint of Linde's article "Particle Physics and Inflationary Cosmology," an original essay about inflation and the multiverse by cosmologist Craig Hogan, and an introduction by curator Craig Burnett, while the Madrid catalog includes an essay about the sculpture by exhibition curator Lynne Cooke, an essay about the accompanying film (also titled *Island Universe*) by film historian Eric de Bruyn, and an essay about the film's musical score by Josiah and the composer Paul Schütze.

Finally, the best source for images of *An End to Modernity* and its companion film *Conceptual Drawings for a Chandelier (1965)* is the catalog published in 2006 by the Wexner Center, titled *Josiah McElheny: Notes for a Sculpture and a Film* and edited by Josiah McElheny and Helen Molesworth. This catalog includes several essays about *An End to Modernity,* some of which reappear in the Skira/Rizzoli book.

More images and links can be found at
`http://www.astronomy.ohio-state.edu/~dhw/McElheny`

### *Acknowledgments*

I am grateful to Jim Phelan for inviting me to participate in this unusual and inspiring symposium, to Helen Molesworth for getting this ball rolling many years ago, to Lisa Florman for teaching me most of what I know about art, and, of course, to Josiah. My work on this project and this article was supported in part by the National Science Foundation and by an AMIAS membership at the Institute for Advanced Study.

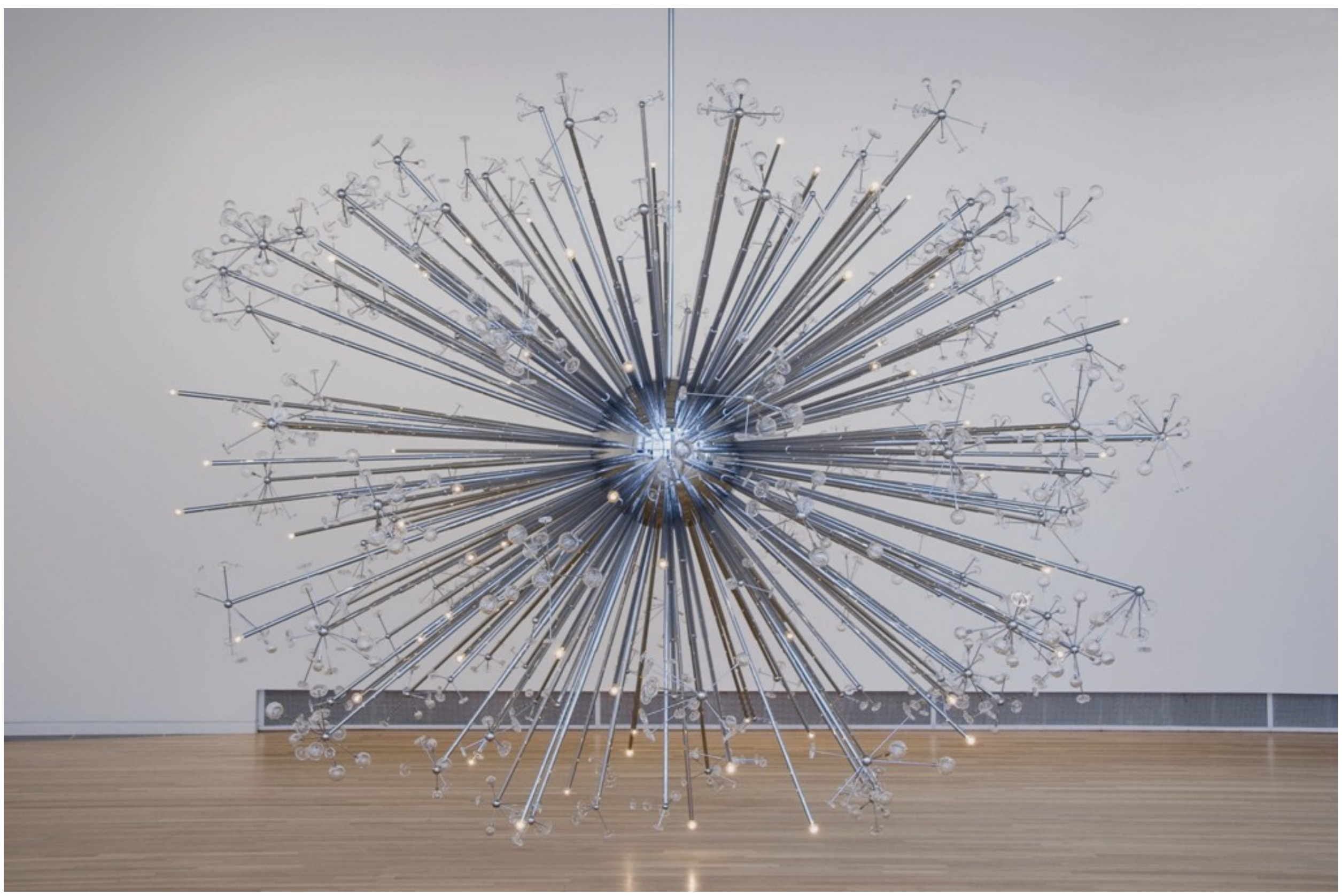

*An End to Modernity* (Josiah McElheny, 2005), exhibited at the Wexner Center for the Arts, Ohio State University, Columbus, Ohio. *An End to Modernity* is now in the permanent collection of the Tate Modern, London. Image courtesy of J. McElheny.

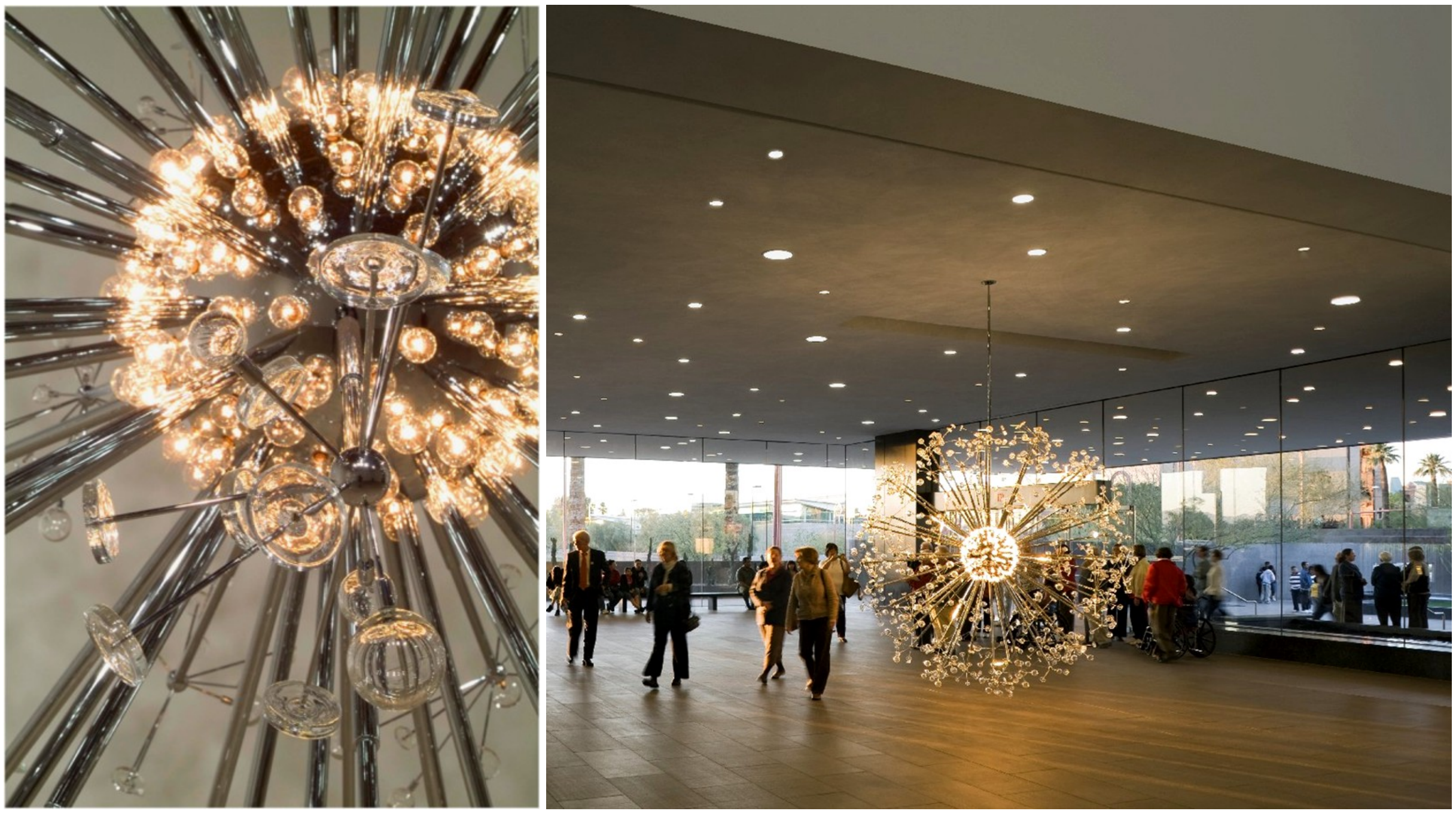

*The Last Scattering Surface* (Josiah McElheny, 2006), exhibited at the Donald Young Gallery in Chicago (left, detail) and in its permanent installation at the Phoenix Art Museum (right). Images courtesy of J. McElheny.

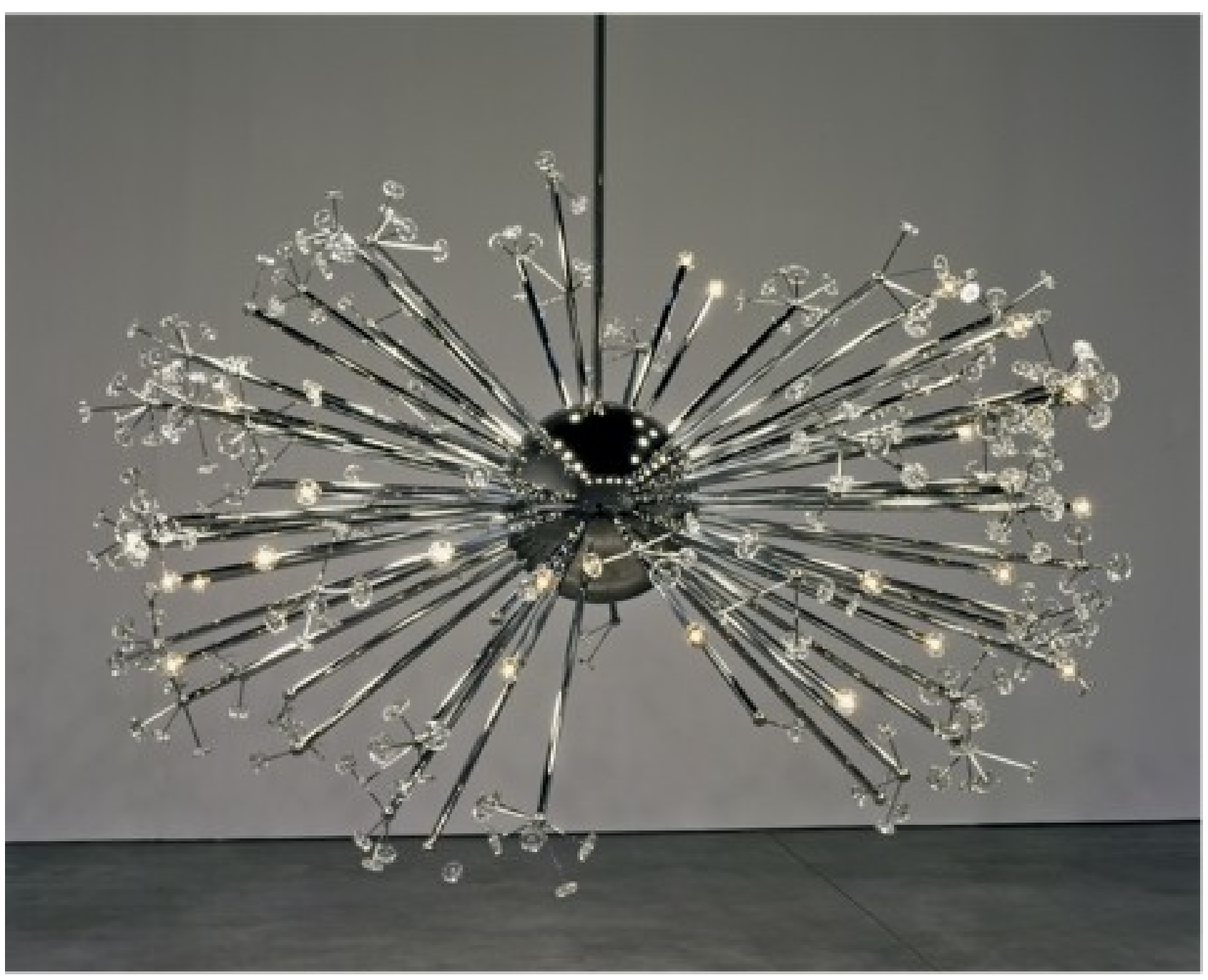

*The End of the Dark Ages* (Josiah McElheny, 2008), exhibited at the Andrea Rosen Gallery in New York City. *The End of the Dark Ages* is now in a private collection. Image courtesy of J. McElheny.

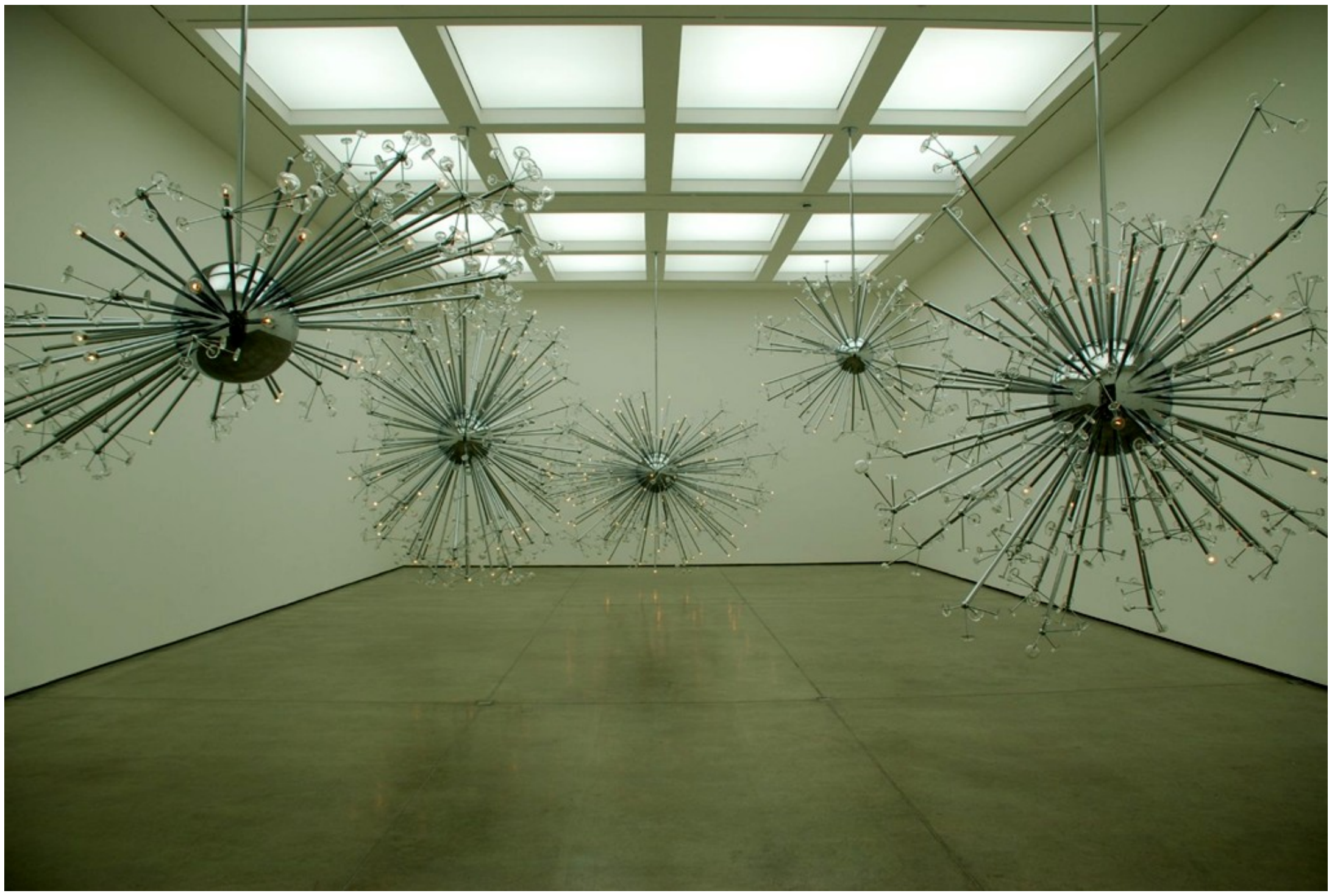

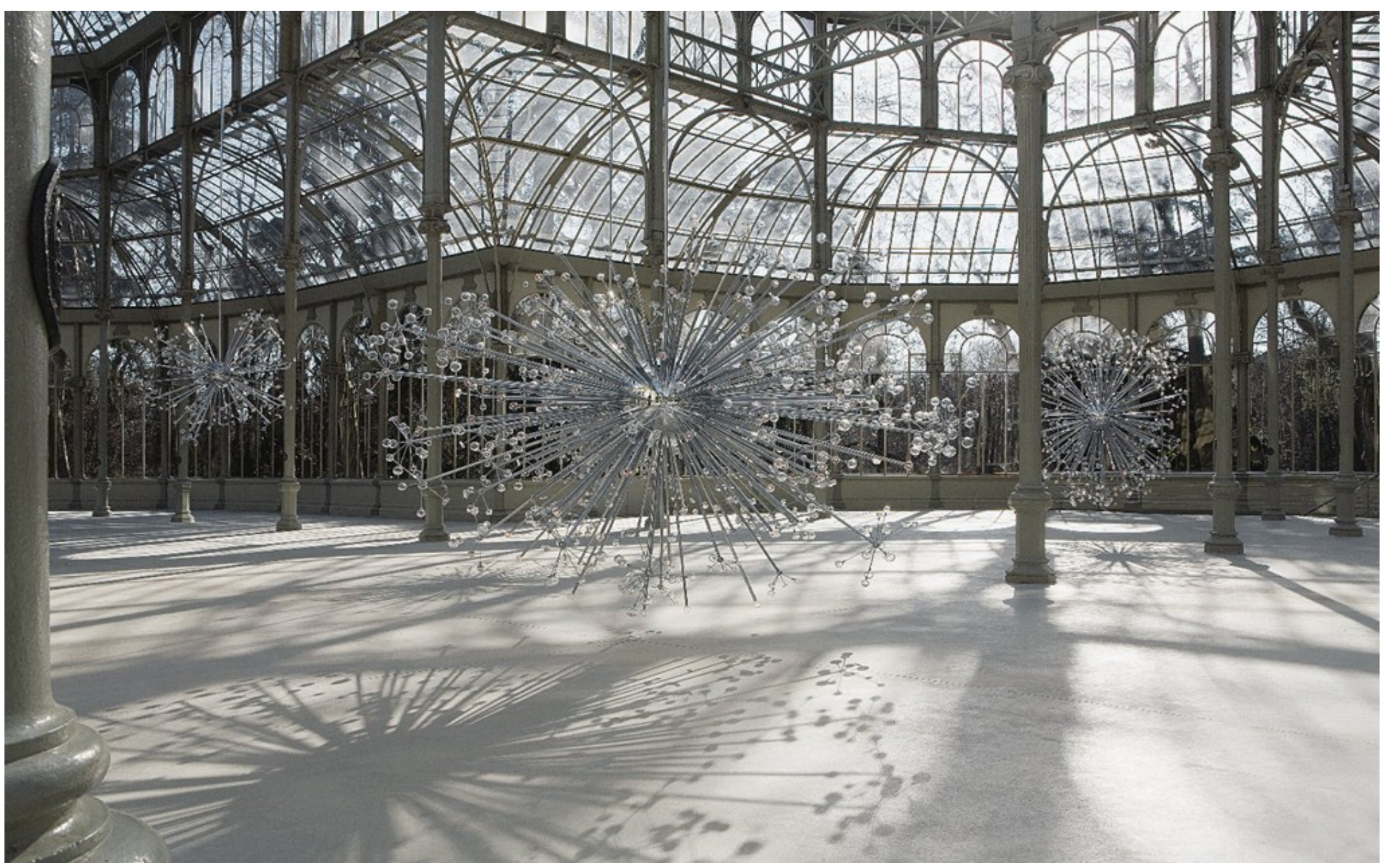

*Island Universe* (Josiah McElheny, 2008), exhibited at the White Cube Gallery, London (top), and at the Palacio de Cristal of the Reina Sofia Museum, Madrid (bottom). Images courtesy of J. McElheny.